\documentclass[aps,pre,amsmath,amssymb,reprint,superscriptaddress]{revtex4-1}
\usepackage[dvipdfmx]{graphicx}
\usepackage{epsf}
\usepackage{bm}
\usepackage{lipsum}
\usepackage[normalem]{ulem}
\def\gs{\gtrsim}
\def\ls{\lesssim}

\usepackage{color}

\begin{document}
\title{Boson peak, elasticity, and glass transition temperature in
polymer glasses:\\ Effects of the rigidity of chain bending}

\author{Naoya Tomoshige}
\affiliation{Division of Chemical Engineering, Department of Materials Engineering Science, Graduate School of Engineering Science, Osaka University, Toyonaka, Osaka 560-8531, Japan}

\author{Hideyuki Mizuno}
\email{hideyuki.mizuno@phys.c.u-tokyo.ac.jp}
\affiliation{Graduate School of Arts and Sciences, The University of Tokyo, Tokyo 153-8902, Japan}

\author{Tatsuya Mori}
\email{mori@ims.tsukuba.ac.jp}
\affiliation{Division of Materials Science, University of Tsukuba, 1-1-1 Tennodai, Tsukuba, Ibaraki 305-8573, Japan}

\author{Kang Kim}
\email{kk@cheng.es.osaka-u.ac.jp}
\affiliation{Division of Chemical Engineering, Department of Materials Engineering Science, Graduate School of Engineering Science, Osaka University, Toyonaka, Osaka 560-8531, Japan}

\author{Nobuyuki Matubayasi}
\email{nobuyuki@cheng.es.osaka-u.ac.jp}
\affiliation{Division of Chemical Engineering, Department of Materials Engineering Science, Graduate School of Engineering Science, Osaka University, Toyonaka, Osaka 560-8531, Japan}
\affiliation{Elements Strategy Initiative for Catalysts and Batteries, Kyoto University, Katsura, Kyoto 615-8520, Japan}

\date{\today}

\begin{abstract}
The excess
 low-frequency vibrational spectrum, called boson peak, and non-affine
 elastic response are the most important particularities of glasses. 
Herein, 
 the vibrational and mechanical properties of polymeric glasses are
 examined by using 
 coarse-grained molecular dynamics simulations, with particular
 attention to the effects of the bending rigidity of the polymer chains. As
 the rigidity increases, the system undergoes a glass transition at a
 higher temperature (under a constant pressure), which decreases the
 density of the glass phase. The elastic moduli, which are
 controlled by the decrease of the density and the increase of
 the rigidity, show a non-monotonic dependence on the rigidity of the polymer
 chain that arises from the non-affine component.
 Moreover, a clear boson peak is observed in the vibrational
 density of states, which depends on the macroscopic shear modulus
 $G$. In particular, the boson peak frequency is scaled as
 $\omega_\mathrm{BP} \propto \sqrt{G}$. These results provide a positive
 correlation between the boson peak, shear elasticity, and the glass
 transition temperature.
\end{abstract}

\maketitle

\section{Introduction}
Glasses show vibrational and mechanical properties that are markedly
different from other crystalline
materials~\cite{Phillips_1981,Alexander_1998}.
Thermal measurements and scattering experiments have been
performed to study the properties of various glassy systems, such as
covalent-bonding~\cite{Zeller_1971,Buchenau_1984,Nakayama_2002,Monaco_2006,Baldi_2010,Chumakov_2011},
molecular~\cite{Yamamuro_1996,Ramos_2003,Monaco_2009,Shibata_2015,Kabeya_2016},
metallic~\cite{Berg_1983,Yong_2008,Bruna_2011,Huang_2014}, and
polymeric~\cite{Niss_2007,Hong_2008,Caponi_2011,Perez-Castaneda2_2014,Terao_2018,Zorn_2018}
glasses.
For instance, the excess vibrational modes at low frequencies and the
excess heat capacity at low temperatures exceeding the Debye predictions, which
describe the corresponding crystalline values, have been 
observed universally in various glassy materials.
This phenomenon, which is referred to as the boson peak~(BP), has been
widely studied.

The ideas of elastic
heterogeneities~\cite{Schirmacher_2006,Schirmacher_2007,Schirmacher_2015}
and criticality near isostatic state and marginally stable
state~\cite{Wyart_2005,Wyart_2006,Wyart_2010,DeGiuli_2014} have been
introduced, following the recent theoretical advances for understanding the origin of anomalies in glasses.
Based on these theories, the mean-field formulations have been developed
by using the effective medium
technique~\cite{Schirmacher_2006,Schirmacher_2007,Schirmacher_2015,Wyart_2010,DeGiuli_2014}.
In addition, more recent studies~\cite{Milkus_2016,Krausser_2017} have
focused on the local inversion-symmetry breaking, which can explain the microscopic origin of the BP.

Molecular dynamics (MD) simulations play an essential role for studying
the vibrational and mechanical properties of glasses.
Firstly, MD simulations enable to assess the theoretical predictions.
In fact, various MD simulations have been performed on simple atomic
glasses, e.g., Lennard-Jones~(LJ)
systems~\cite{Schober_1996,Mazzacurati_1996,Shintani_2008,Monaco2_2009,Mizuno2_2013,Lerner_2016,Wang_2019}.
Concerning the isostaticity and marginal
stability~\cite{Wyart_2005,Wyart_2006,Wyart_2010,DeGiuli_2014}, the
systems with a finite-ranged, purely repulsive potential have also been
studied~\cite{Silbert_2005,Silbert_2009,Vitelli_2010,Xu_2010}, and are considered as the simplest model of glasses.
In particular, it is crucial for MD simulations to solve finite-dimensional
effects that are not captured by the mean-field
treatments~\cite{Mizuno_2017,Shimada_2018,Mizuno_2018}.
Secondly, MD simulations perform quasi-experiments on well-defined
systems and access data that cannot be examined experimentally.
Relevant systems to experiments and applications have been simulated,
including
covalent-bonding~\cite{Taraskin_1997,Taraskin_1999,Horbach_2001,Leonforte_2006,Beltukov_2016,Beltukov_2018},
metallic~\cite{Derlet_2012,Fan_2014,Crespo_2016,Brink_2016},
polymeric~\cite{Jain_2004,Schnell:2011he,Ness:2017cc,Milkus:2018ha,Giuntoli:2018fv}
glasses.
These simulation studies complete theoretical understandings based on
simple systems and experimental observations of more complex systems.

The vibrational properties and the BP of polymeric glasses have been
studied by both of
experiments~\cite{Niss_2007,Hong_2008,Caponi_2011,Perez-Castaneda2_2014,Terao_2018,Zorn_2018}
and MD
simulations~\cite{Jain_2004,Schnell:2011he,Ness:2017cc,Milkus:2018ha,Giuntoli:2018fv}.
The effects caused by 
non-covalent bonds including bending forces and chain length represent an important feature of polymer glasses.
Previous experiments~\cite{Niss_2007,Hong_2008} have investigated the
effects of the pressure or densification on the frequency and intensity of the BP in polymeric glasses.
It was demonstrated that the evolution of the BP with pressure cannot be
scaled by the Debye values~(i.e., the Debye frequency and the Debye
level).
Therefore, the pressure effects cannot be explained only by the variation of macroscopic elasticity.
In contrast, another experiment~\cite{Caponi_2011} has shown that the
polymerization effects on the BP is explained by the change in
macroscopic elasticity as the frequency and intensity variations of the BP are both scaled by the Debye values.

In addition, Zaccone \textit{et al.} have recently performed MD
simulations to calculate the vibrational density of states~(vDOS) in
polymeric glasses by changing the chain length and the rigidity of the chain bending~\cite{Milkus:2018ha}.
This work studied the vibrational eigenstates in a wide range of
frequencies and the effects of the chain length and bending rigidity on
the high-frequency spectra.
Furthermore, Giuntoli and Leporini studied the BP of polymeric glasses
having chains with highly rigid bonds~\cite{Giuntoli:2018fv}.
It was demonstrated that the BP decouples with macroscopic elasticity
and arises from non-bonding interactions only.
Although these studies~\cite{Milkus:2018ha,Giuntoli:2018fv} have helped
understand polymeric glass properties, the effects of bending rigidity
and chain length on the low-frequency spectra and BP need to be further studied.

Herein, the vDOS and the elastic moduli of polymeric glasses are
analyzed through coarse-grained MD simulations.
In particular, the connection between the BP and elasticity as well as
the glass transition temperature is explored by systematically changing
the bending stiffness of short and long polymer chains.
The contributions of the present study are given as follows.
We demonstrate that polymeric glasses can exhibit extremely-large
non-affine elastic response~(compared to atomic glasses), whereas the BP
is simply scaled by the behavior of macroscopic shear modulus.
This behavior of the BP can be explained by the theory of elastic
heterogeneities~\cite{Schirmacher_2006,Schirmacher_2007,Schirmacher_2015}.
Our results indicate that effects of the bending rigidity on the BP are
encompassed in change of macroscopic elasticity, which is in contrast to
effects of pressure~\cite{Niss_2007,Hong_2008}, but instead is similar
to effects of polymerization~\cite{Caponi_2011}.
Furthermore, we show the positive correlation among the BP, elasticity,
and the glass transition temperature.
Finally, we will discuss the relaxation dynamics in the liquid state, in
relation to our results of low-frequency vibrational spectra.

\section{System Description}
Coarse-grained MD simulations are performed by using the Kremer--Grest
model~\cite{Kremer:1990iv}, which treats polymer chains as linear series
of monomer beads (particles) of mass $m$.
Each polymer chain is composed of $L$ monomer beads, and two cases are
considered in this study: long chain length with $L=50$ and short chain
length with $L=3$.
In a three-dimensional cubic simulation box under periodic boundary
conditions, $N_\mathrm{p} = 5000$ and $4998$ is defined as the total
number of 
monomers for $L=50$ and $L=3$ respectively, which means that the number
of polymeric chains is $N_\mathrm{p}/L = 100$ for $L=50$ and $1666$ for $L=3$.

The polymer chain is modeled by three types of inter-particle potentials as follows.
Firstly, all the monomer particles interact via the LJ potential:
\begin{equation}
U_\mathrm{LJ}(r) = 4\varepsilon_\mathrm{LJ} \left[ \left( \frac{\sigma}{r}
					    \right)^{12} - \left(
					    \frac{\sigma}{r} \right)^{6}
				     \right],
\label{eq:LJ}
\end{equation}
where $r$ is the distance between two monomers, $\sigma$ is the diameter
of monomer, and $\varepsilon_\mathrm{LJ}$ is the energy scale of the LJ
potential.
The LJ potential is truncated at the cut-off distance of $r_c = 2.5
\sigma$, where the potential and the force (first derivative of the
potential) are shifted to zero continuously~\cite{Shimada:2018fp}.
Throughout this study, the mass, length, and energy scales are measured
in units of $m$, $\sigma$, $\varepsilon_\mathrm{LJ}$, respectively.
The temperature is measured by
$\varepsilon_\mathrm{LJ}/k_\mathrm{B}$~($k_\mathrm{B}$ is the Boltzmann
constant). 
Secondly, sequential monomer-beads along the polymeric chain are
connected by a finitely extensible nonlinear elastic~(FENE) potential:
\begin{equation}
U_\mathrm{FENE}(r) =
\left\{ 
\begin{aligned}
& -\frac{\varepsilon_\mathrm{FENE}}{2} R_0^2 \ln \left[ 1 - \left(
 \frac{r}{R_0} \right)^{2} \right] & (r \le R_0), \\
& \infty & (r > R_0),
\end{aligned} 
\right.
\label{eq:FENE}
\end{equation}
where $\varepsilon_\mathrm{FENE}$ is the energy scale of the FENE
potential, and $R_0$ is the maximum length of the FENE bond.
Their values are defined as 
$\varepsilon_\mathrm{FENE} = 30$ and $R_0 = 1.5$, according to Ref.~\cite{Milkus:2018ha}.
Finally, three consecutive monomer beads along the chain interact via
the bending potential defined as follows:
\begin{equation}
U_\mathrm{bend}(\theta) = \varepsilon_\mathrm{bend} \left[ 1 - \cos(\theta -
					      \theta_0) \right],
\label{eq:bend}
\end{equation}
where $\theta$ is the angle formed by three consecutive beads, and
$\varepsilon_\mathrm{bend}$ is the associated energy scale.
This potential intends to stabilize the angle $\theta$ at $\theta_0$
that we set as $\theta_0 = 109.5^\circ$.
Here, the value of $\varepsilon_\mathrm{bend}$ in a wide range from
$\varepsilon_\mathrm{bend} = 10^{-3}$ to $10^4$, and the effects of the
bending rigidity on the vibrational and mechanical properties of the
polymeric system are studied.

MD simulations are performed by using the LAMMPS~\cite{Plimpton_1995,lammps}.
The polymeric system is first equilibrated in the melted, liquid state
at a temperature $T=1.0$.
Further, 
the system is cooled down under a fixed pressure condition of {$P =0$}
and with a cooling rate of $dT/dt = 10^{-4}$.
During the cooling process, the glass transition occurs at a particular
temperature, i.e., the glass transition temperature.
After the glass transition, the system is quenched down towards the zero
temperature, i.e., $T=0$ state.

\begin{figure}[t]
\centering
\includegraphics[width=0.48\textwidth]{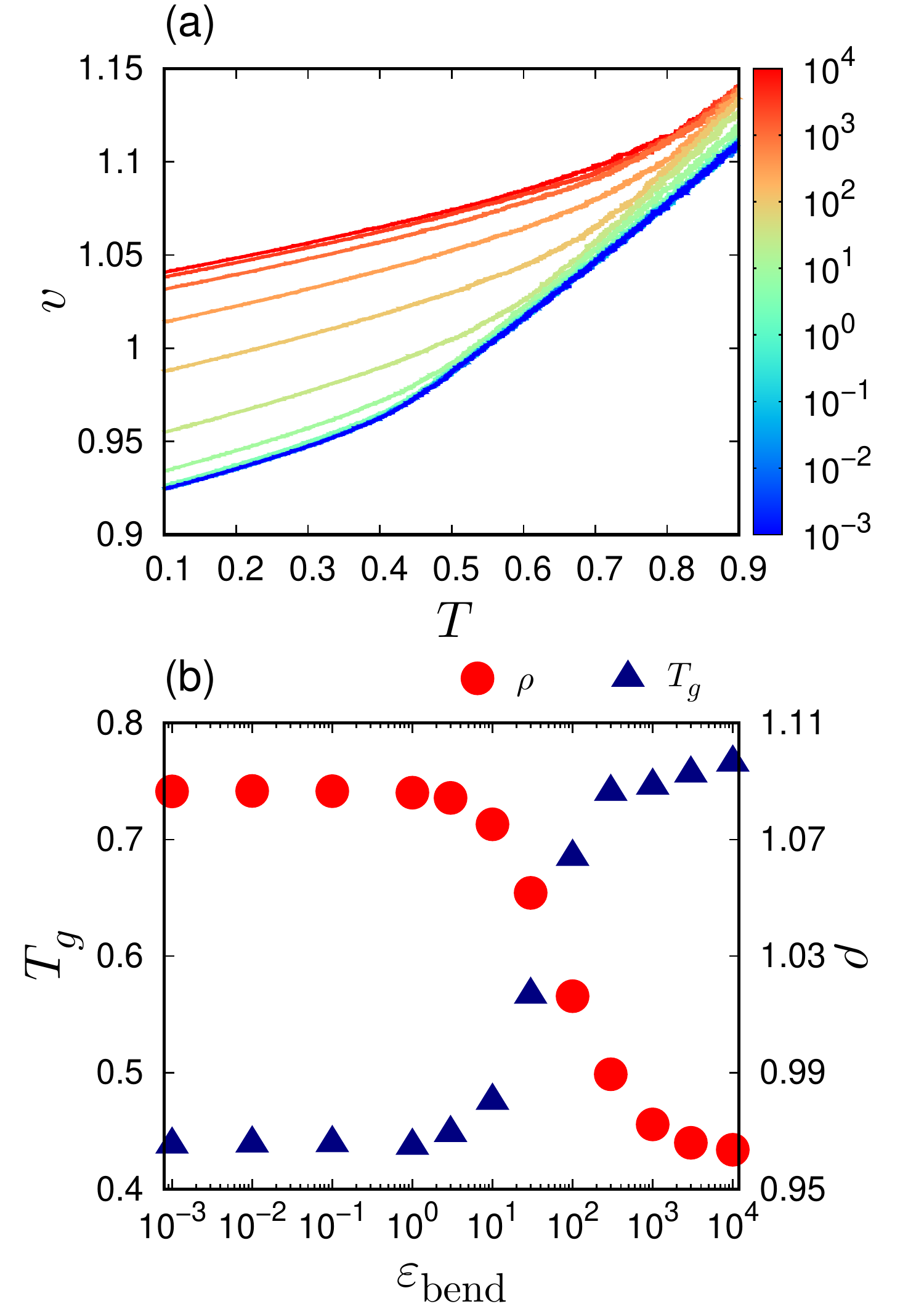}
\vspace*{0mm}
\caption{\label{fig1}
{Glass transition temperature and density in the glass state.}
(a) The specific volume $v$ versus the temperature $T$ during the process that the system is cooled down from the liquid state to the glass state.
The color of line indicates the value of bending rigidity $\varepsilon_\mathrm{bend}$ according to the color bar.
(b) Glass transition temperature $T_g$ (triangles) and  density $\rho$ at zero temperature after the glass transition (circles) are plotted against $\varepsilon_\mathrm{bend}$.
The chain length is $L=50$.
}
\end{figure}

\begin{figure}[t]
\centering
\includegraphics[width=0.48\textwidth]{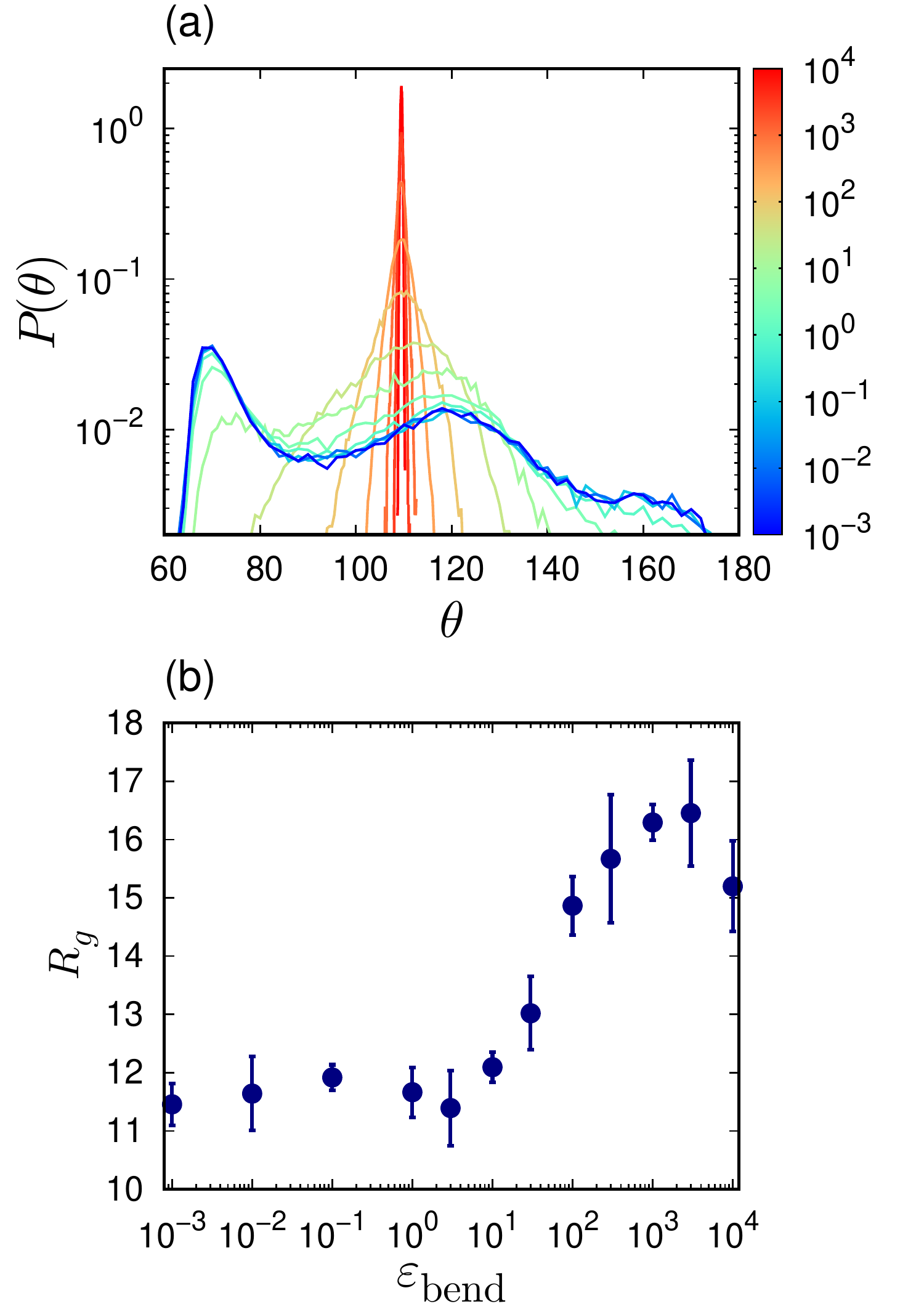}
\vspace*{0mm}
\caption{\label{fig2}
{Conformation of polymeric chains.}
(a) Probability distribution of angle formed by three consecutive beads along the chain, $P(\theta)$, is presented for several different rigidities $\varepsilon_\mathrm{bend}$.
The color of line indicates the value of bending rigidity $\varepsilon_\mathrm{bend}$ according to the color bar.
(b) Radius of inertia $R_g$ is plotted as a function of $\varepsilon_\mathrm{bend}$.
The chain length is $L=50$.
}
\end{figure}

\begin{figure*}[t]
\centering
\includegraphics[width=0.96\textwidth]{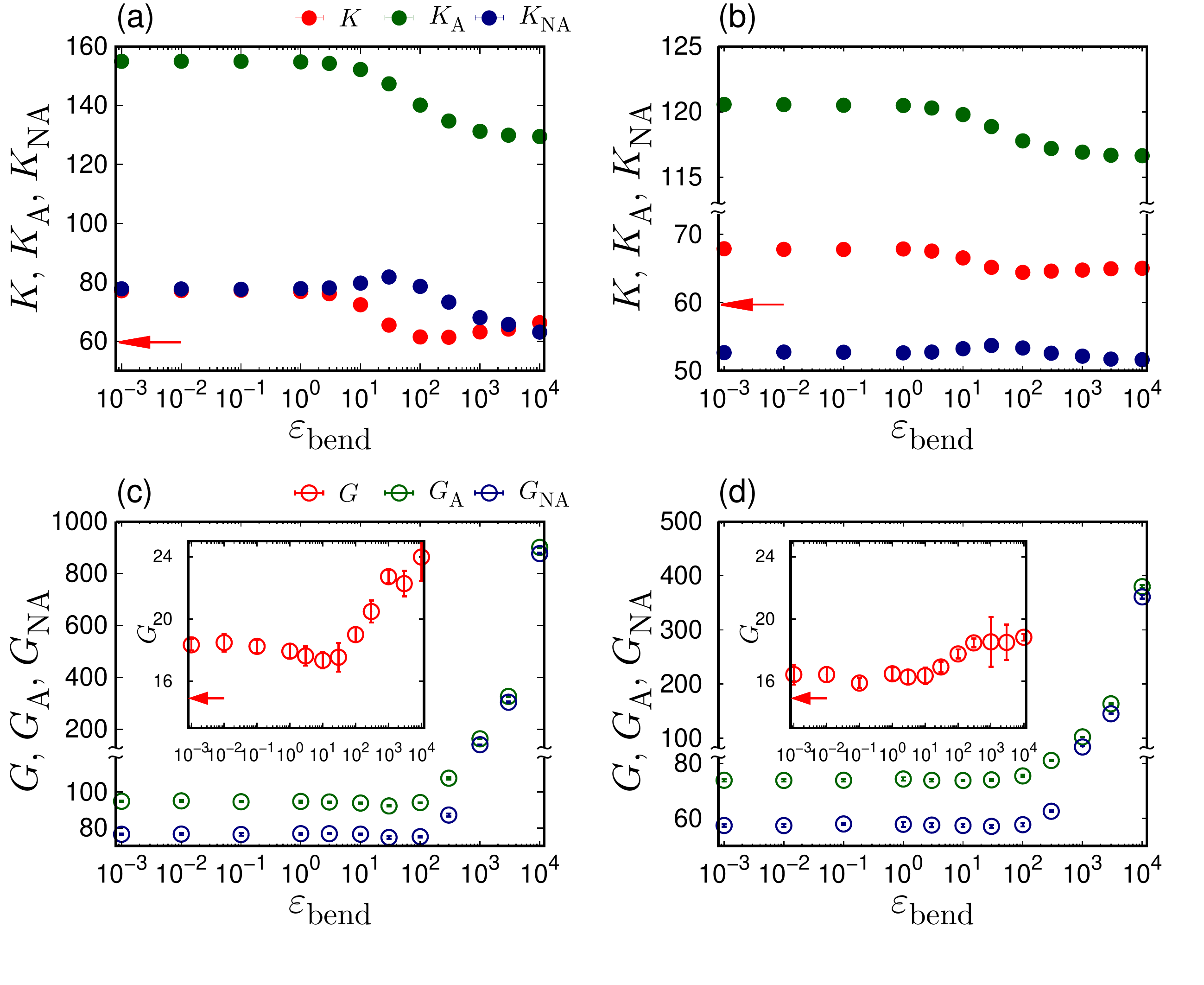}
\vspace*{0mm}
\caption{\label{fig3}
{Elastic properties of polymeric glasses.}
Plots of the bulk modulus $K$ [(a),(b) upper panels] and the shear modulus $G$ [(c),(d) bottom panels] as functions of the strength of bending rigidity $\varepsilon_\mathrm{bend}$.
The chain length is $L=50$ [(a),(c) left panels] and $L=3$ [(b),(d) right panels].
In the figures, we also plot the affine moduli, $K_\mathrm{A}$ and $G_\mathrm{A}$, and the non-affine moduli, $K_\mathrm{NA}$ and $G_\mathrm{NA}$.
The horizontal arrows indicate the values, $K = 59.7$ and $G = 14.9$, of
 atomic LJ glasses that are extracted from Ref.~\cite{Mizuno_2013}.
}
\end{figure*}

\section{Results}

\subsection{Glass transition temperature}
When the polymeric system is cooled down from the liquid state under a
constant pressure, the volume of the system monotonically decreases with
decreasing the temperature.
Figure~\ref{fig1}a shows the specific volume $v$ as a function of the
temperature $T$ for several different bending rigidities
$\varepsilon_\mathrm{bend}$ and the chain length $L=50$.
For each value of $\varepsilon_\mathrm{bend}$, the slope of the $v$-$T$
curve clearly presents a discontinuous change at a certain temperature,
which is defined as the glass transition temperature $T_g$.
Figure~\ref{fig1}b~(triangles) presents the value of $T_g$ as a function of $\varepsilon_\mathrm{bend}$.
As the rigidity increases from $\varepsilon_\mathrm{bend} = 1$ to
$10^3$, $T_g$ progressively increases from $T_g \simeq 0.45$ to $0.75$.
Below $\varepsilon = 1$ and above $\varepsilon = 10^3$, the variation of $T_g$ is low or even negligible.
In addition, Figure~\ref{fig1}b~(circles) plots the density $\rho(=1/v)$
of the system that is quenched down to $T=0$.
The density decreases from $\rho \simeq 1.09$ to $0.97$ as the rigidity
increases from $\varepsilon_\mathrm{bend} =1$ to $10^3$.
As the chain bending becomes rigid, the glass transition occurs at a
higher temperature, and as a result, the density in the glass state
becomes lower.
The similar observation was obtained by Milkus \textit{et al}~\cite{Milkus:2018ha}.

These behaviors of $T_g$ and $\rho$ can be understood by studying the
microscopic conformation of the polymeric chains.
Figure~\ref{fig2}a presents the probability distribution of the angle
formed by three consecutive beads along the chain, $P(\theta)$, when
changing the rigidity $\varepsilon_\mathrm{bend}$.
Two peaks are observed at approximately $\theta \simeq 70^\circ$ and
$\theta \simeq 120^\circ$ for a low rigidity ($\varepsilon_\mathrm{bend}\le 1$).
A similar distribution $P(\theta)$ was also reported in Ref.~\cite{Milkus:2018ha}.
As the rigidity increases, the peak position in $P(\theta)$ shifts towards $\theta_0=109.5^\circ$.
It is noted that the bending potential $U_\mathrm{bend}(\theta)$ in
Eq.~(\ref{eq:bend}) tends to stabilize the angle $\theta$ at $\theta_0 =
109.5^\circ$.
In addition, Figure~\ref{fig2}b presents the radius of gyration $R_g$ as
a function of $\varepsilon_\mathrm{bend}$.
It can be observed that $R_g$ increases from $R_g \simeq 11.5$ to $16.5$
with an increasing $\varepsilon_\mathrm{bend}$.
Importantly, these variations of conformation are induced intensively
when the rigidity increases from $\varepsilon_\mathrm{bend} = 1$ to
$10^3$, which exactly matches the region where variations of $T_g$ and
$\rho$ are observed in Fig.~\ref{fig1}b.
Therefore, it can be concluded that the conformation changes of the polymeric chains
control the glass transition temperature and the density.
In fact, as the
rigidity of the chain bending increases, the angle $\theta$ of the polymer
chains tends to be stabilized at $\theta_0 = 109.5^\circ$ and the radius
of inertia increases.
As a result, the glass transition occurs
at a higher temperature and the lower density (larger volume).
At $\varepsilon_\mathrm{bend} \ls 1$, the effect of the
bending interaction of Eq.~(\ref{eq:bend}) is weak compared to those of the
LJ and FENE components of Eqs.~(\ref{eq:LJ}) and (\ref{eq:FENE}).
However, at $\varepsilon_\mathrm{bend} \gs 10^3$,
the opposite phenomenon occurs.

It is noted that the glass transition occurs at a lower temperature for
$L=3$ than for $L=50$, which is consistent with a previous report~\cite{Durand_2010}.
Correspondingly, the values of $\rho$ for $L=3$ becomes larger than that of $L=50$.
However, common results were observed between $L=3$ and $50$ with
respect to the dependences on the rigidity $\varepsilon_\mathrm{bend}$.
Specifically, $T_g$ and $\rho$, as well as the conformation of the polymeric
chains progressively change when the rigidity increases from
$\varepsilon_\mathrm{bend} = 1$ to $10^3$, which also occurs for 
$L=50$.

\subsection{Elastic properties}
The elastic properties of polymer glasses are studied by changing the strength of bending rigidity.
An external strain is applied to the system at $T=0$, which enables to
measure the corresponding elastic moduli.
Specifically, the volume-changing bulk deformation and the
volume-conserving shear deformation are applied, which provide the bulk
modulus $K$ and the shear modulus $G$, respectively~\cite{Mizuno_2013}.
Figure~\ref{fig3} presents the values of $K$ and $G$ as functions of $\varepsilon_\mathrm{bend}$.
Disordered systems exhibit large non-affine elastic
responses~\cite{Alexander_1998}.
The elastic moduli, $M=K$ and $G$, are decomposed into affine moduli
$M_\mathrm{A}$ and non-affine moduli $M_\mathrm{NA}$, i.e., $M =
M_\mathrm{A} - M_\mathrm{NA}$~\cite{Tanguy_2002,Lemaitre_2006}.
In Fig.~\ref{fig3}, these affine and non-affine components are also presented.

First, the bulk modulus $K$ is analyzed for $L=50$ and presented Fig.~\ref{fig3}a.
The affine component $K_\mathrm{A}$ decreases from $K_\mathrm{A} \simeq
155$ to $130$ as $\varepsilon_\mathrm{bend}$ changes from $1$ to $10^3$.
The reduction of $K_\mathrm{A}$ is caused by the decrease of the density
$\rho$ with the increasing $\varepsilon_\mathrm{bend}$ (see Fig.~\ref{fig1}b).
In contrast, the non-affine component $K_\mathrm{NA}$ shows a
non-monotonic dependence on the $\varepsilon_\mathrm{bend}$.
In particular, $K_\mathrm{NA}$ slightly increases from
$\varepsilon_\mathrm{bend} = 1$ to $30$, which is induced by the
decrease of the density $\rho$.
As $\varepsilon_\mathrm{bend}$ is further increased above $\varepsilon_\mathrm{bend} = 30$, $K_\mathrm{NA}$ decreases.
This is because the non-affine relaxation process is constrained due to
the large rigidity of $\varepsilon_\mathrm{bend}$.
As a result, the total modulus of $K=K_\mathrm{A}-K_\mathrm{NA}$ also
presents a non-monotonic behavior, which is demonstrated in
Fig.~\ref{fig3}a.
From $\varepsilon_\mathrm{bend}=1$ to $10^2$, $K$ decreases from $K
\simeq 80$ to $60$, which is caused by the reduction of $K_\mathrm{A}$.
Moreover, $K$ increases from $K \simeq 60$ to $65$ above
$\varepsilon_\mathrm{bend} = 100$, which is caused by the reduction of
$K_\mathrm{NA}$.
Therefore,
the $\varepsilon_\mathrm{bend}$ dependence of the bulk modulus $K$ is
determined by the competition between the density reduction and the
increase in the bending rigidity.

Further, the shear modulus $G$ is analyzed for $L=50$ and presented Fig.~\ref{fig3}c.
It can be observed that the bending rigidity strongly affects the shear
modulus compared to the bulk modulus.
Particularly, above $\varepsilon_\text{bend} = 10^2$, both of the affine
$G_\mathrm{A}$ and non-affine $G_\mathrm{NA}$ components considerably
increase.
As the shear deformation is anisotropic and causes deformations of
the angles $\theta$ of polymeric chains, its response is expected to be
highly affected by the bending rigidity.
Interestingly, contrary to the important increases of $G_\mathrm{A}$ and
$G_\mathrm{NA}$, the total shear modulus $G = G_\mathrm{A} -
G_\mathrm{NA}$ shows a low variation (by comparing $G_\mathrm{A} \simeq
G_\mathrm{NA} \simeq 900$ with $G \simeq 24$ at
$\epsilon_\text{bend}=10^4$).
The bending rigidity increases the affine shear modulus but, at the same
time, the non-affine component also increases to cancel the increase in
$G_\mathrm{A}$, and as a result, the total shear modulus presents a low
increase.
The elasticity of the shear deformation is therefore different
from that of the bulk deformation, which is obvious when the elastic
moduli are decomposed into affine and non-affine components.

Figure~\ref{fig3} also shows $K$ in (b) and $G$ in (d) for $L=3$.
The values of $K$ and $G$ of $L=3$ are smaller than those of $L=50$, due
to the bonding energy, $\varepsilon_\mathrm{FENE}$, connecting the
monomers along the polymeric chains.
The responses of $K$ and $G$ to the variation of
$\varepsilon_\mathrm{bend}$ are also weaker for $L = 3$.
However, $K$ and $G$, as well as affine $K_\mathrm{A}$ and
$G_\mathrm{A}$ and non-affine $K_\mathrm{NA}$ and $G_\mathrm{NA}$,
exhibit overall common dependences on $\varepsilon_\mathrm{bend}$
between $L=3$ and $50$.
Therefore, the decrease in $\rho$ and increase in
$\varepsilon_\mathrm{bend}$ engenders similar effects on the elasticity for $L=3$ and $50$.

Finally, it is remarked that the polymer glasses present larger non-affine elastic
components than the atomic (LJ) glasses~\cite{Leonforte_2005,Mizuno_2013}.
Even under an isotropic bulk deformation, the non-affine
$K_\mathrm{NA}$~($\simeq 80$ for $L=50$ and $\simeq 50$ for $L=3$, at
$\varepsilon_\mathrm{bend}\le 1$) is approximately half of the magnitude of the
affine $K_\mathrm{A}$~($\simeq 155$ for $L=50$ and $\simeq 120$ for
$L=3$, at $\varepsilon_\mathrm{bend}\le 1$).
This result is different from that of the LJ glasses, where a negligible
value of $K_\mathrm{NA} \simeq 0.5$~(whereas $K_\mathrm{A} \simeq 60.2$)
was obtained~\cite{Mizuno_2013}.
Larger non-affine moduli reflect various elastic responses due to the
multiple degrees of conformations in polymeric chains.
Therefore, the non-affine deformation process must be considered to
characterize the elastic property of polymeric systems.

\begin{figure}[t]
\centering
\includegraphics[width=0.48\textwidth]{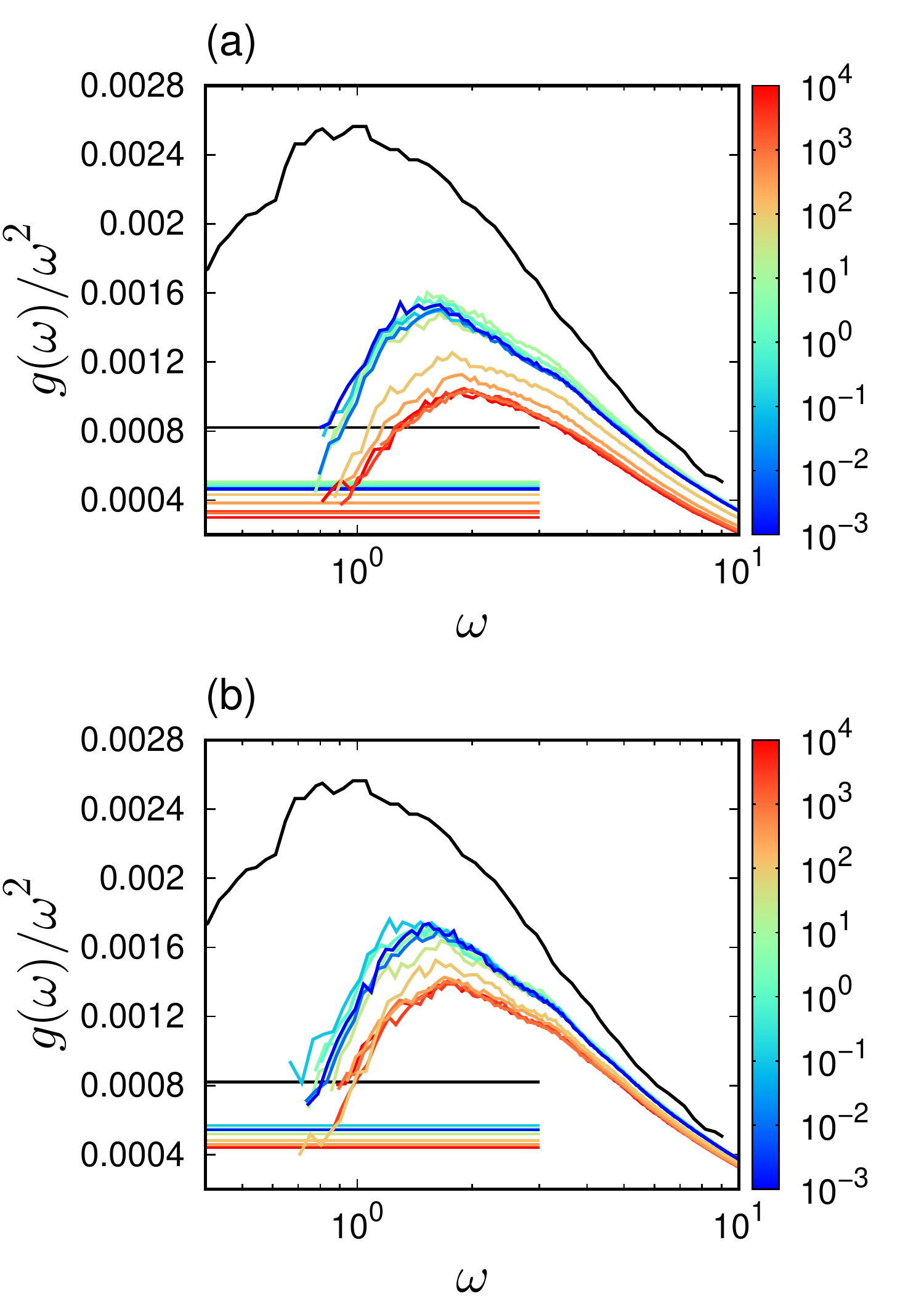}
\vspace*{0mm}
\caption{\label{fig4}
{Low-frequency vibrational spectra.}
We plot the vDOS $g(\omega)$ divided by $\omega^2$, i.e., the reduced vDOS $g(\omega)/\omega^2$, with changing the strength of bending rigidity $\varepsilon_\mathrm{bend}$.
The chain length is (a) $L=50$ and (b) $L=3$.
The horizontal lines indicate the Debye level $A_\mathrm{D}$.
The color of line indicates the value of bending rigidity $\varepsilon_\mathrm{bend}$ according to the color bar.
Black lines present value of the LJ glass which is taken from Ref.~\cite{Shimada:2018fp}.
}
\end{figure}

\begin{figure}[t]
\centering
\includegraphics[width=0.48\textwidth]{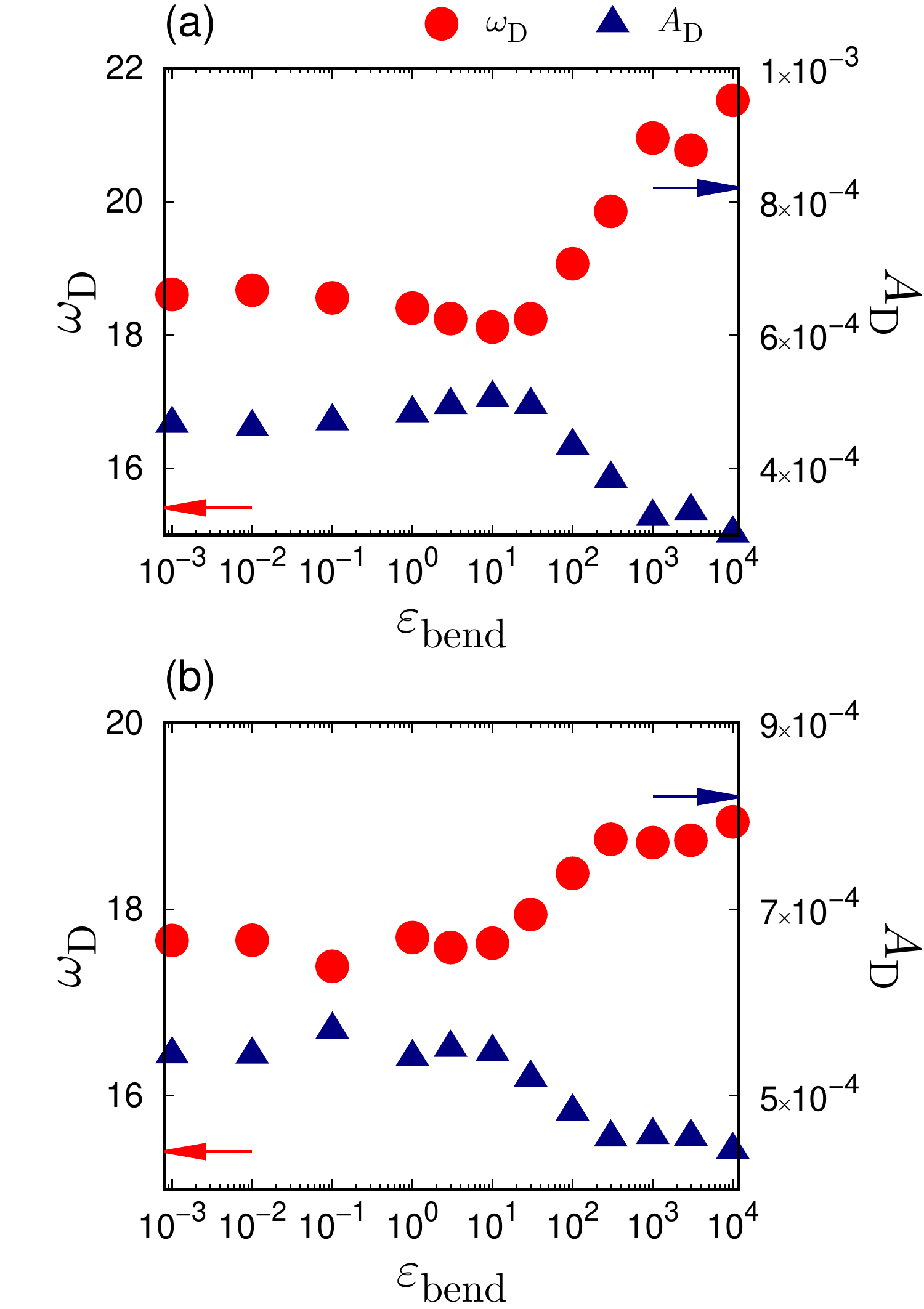}
\vspace*{0mm}
\caption{\label{fig5}
{Debye frequency and Debye level.}
Plots of the Debye frequency $\omega_\mathrm{D}$ (circles) and the Debye level $A_\mathrm{D} = 3/\omega_\mathrm{D}^3$ (triangles) as functions of the strength of bending rigidity $\varepsilon_\mathrm{bend}$.
The chain length is (a) $L=50$ and (b) $L=3$.
The values of $\omega_\mathrm{D}$ and $A_\mathrm{D}$ are calculated from the elastic moduli of $K$ and $G$ that are presented in Fig.~\ref{fig3}.
The arrows indicate values of atomic LJ glasses that are taken from Ref.~\cite{Shimada:2018fp}.
}
\end{figure}

\begin{figure}[t]
\centering
\includegraphics[width=0.48\textwidth]{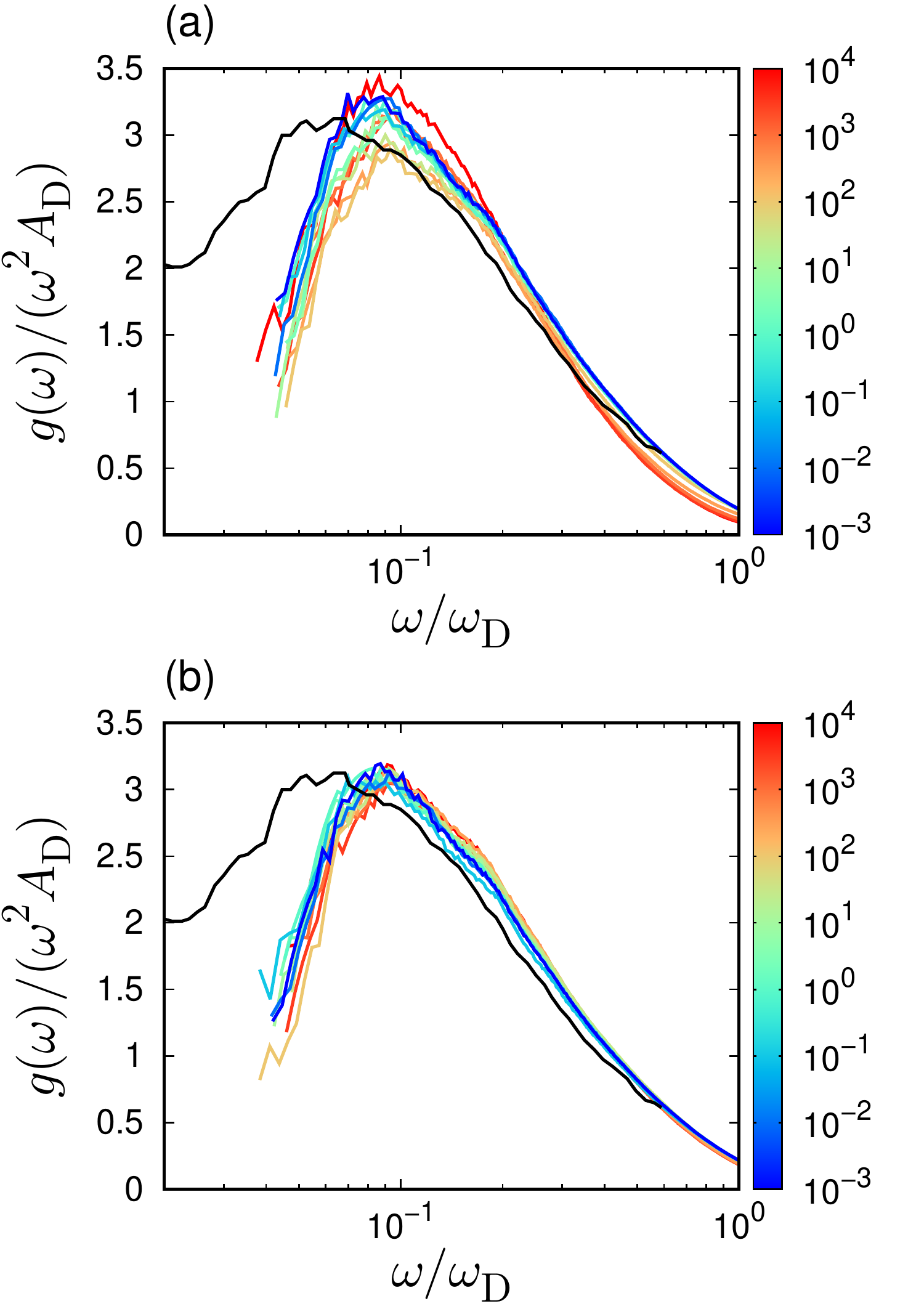}
\vspace*{0mm}
\caption{\label{fig6}
{Scaled vibrational spectra.}
We present the data presented in Fig.~\ref{fig4}, in the scaled form: we scale the reduced vDOS $g(\omega)/\omega^2$ and the frequency $\omega$ by the Debye level $A_\mathrm{D}$ and the Debye frequency $\omega_\mathrm{D}$.
Here the values of $A_\mathrm{D}$ and $\omega_\mathrm{D}$ are presented in Fig.~\ref{fig5}.
The chain length is (a) $L=50$ and (b) $L=3$.
The color of line indicates the value of bending rigidity $\varepsilon_\mathrm{bend}$ according to the color bar.
Black lines present value of the LJ glass which is taken from Ref.~\cite{Shimada:2018fp}.
}
\end{figure}

\begin{figure}[t]
\centering
\includegraphics[width=0.48\textwidth]{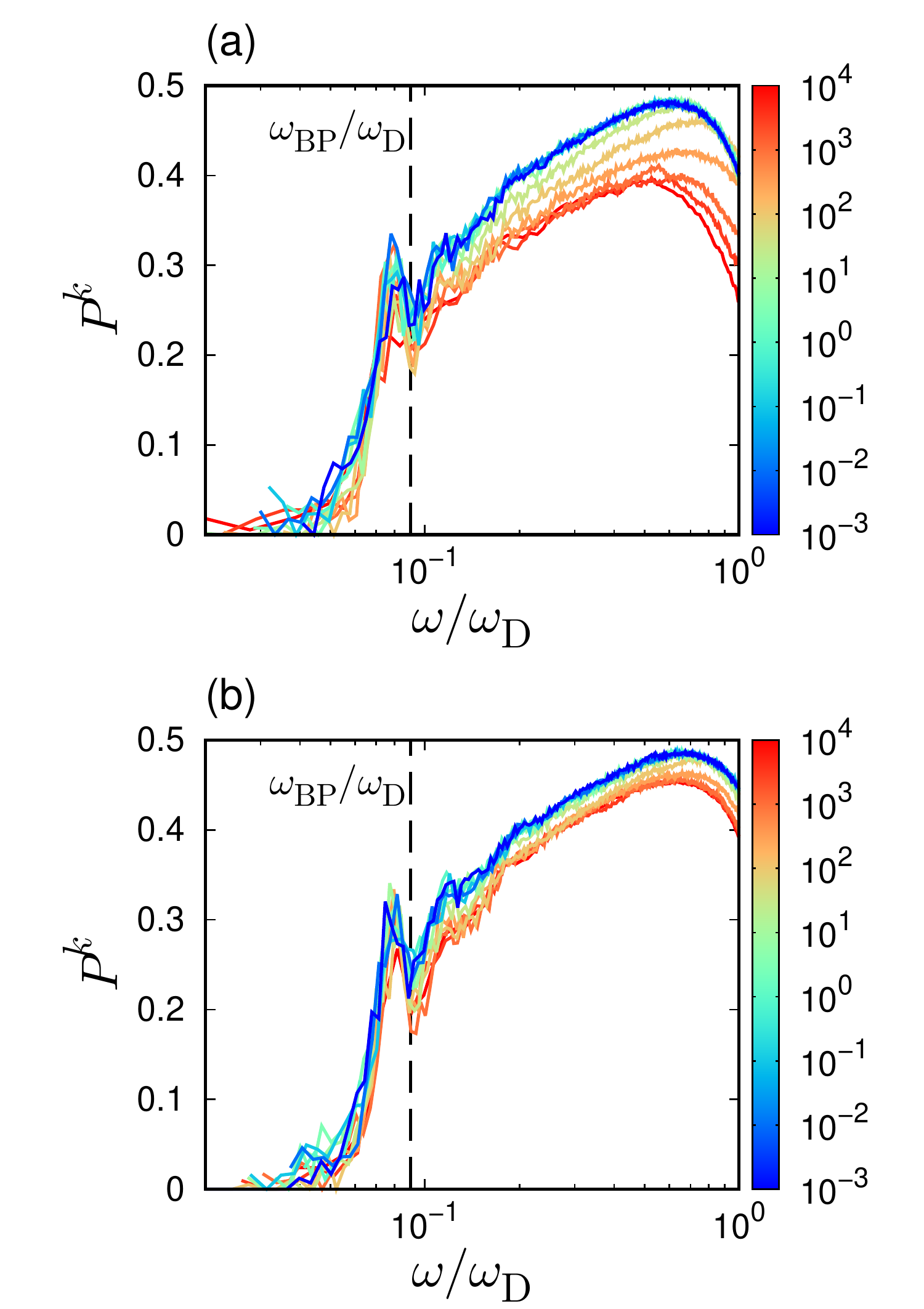}
\vspace*{0mm}
\caption{\label{fig7}
{Localization nature of vibrational states.}
Plots of participation ratio $P^k$ as a function of the scaled frequency $\omega/\omega_\mathrm{D}$, for several different bending rigidities of $\varepsilon_\mathrm{bend}$.
The chain length is (a) $L=50$ and (b) $L=3$.
The color of line indicates the value of bending rigidity $\varepsilon_\mathrm{bend}$ according to the color bar.
Data are shown as the average values over bins in the frequency domain of $\left[ \omega - \Delta\omega/2, \omega + \Delta\omega/2\right]$ with $\Delta \omega \simeq 0.06$.
The vertical line indicates the position of $\omega_\mathrm{BP}/\omega_\mathrm{D}$ averaged over the examined systems with varied $\varepsilon_\mathrm{bend}$.
}
\end{figure}

\subsection{Low-frequency vibrational spectra}
%
\subsubsection{Reduced vDOS}
Finally, the spectra of vibrational eigenmodes in polymer glasses are studied.
The vibrational mode analysis is performed on the configuration of the
polymeric system at $T=0$, which corresponds to the inherent
structure~\cite{Kittel_1996,Ashcroft}.
The Hessian matrix is diagonalized to obtain the eigenfrequencies
$\omega^k$ that corresponds to the square root of the eigenvalues
$\lambda^k$, i.e., $\omega^k = \sqrt{\lambda^k}$~($k=1,2,...,3N_\mathrm{p}$).
The specific expression of the Hessian matrix is given in Supplementary Material
\footnote{
The expression of the Hessian matrix is already described in Ref.~\cite{Milkus:2018ha}.
However, the expression includes errors.
Therefore, the corrected expression is provided in the Supplementary Material.
}.

The statistics of the eigenfrequency provide the vDOS, $g(\omega)$.
Figure~\ref{fig4} presents the reduced version of the vDOS,
$g(\omega)/\omega^2$, when changing the rigidity
$\varepsilon_\mathrm{bend}$ and for $L=50$ in (a) and $L=3$ in (b).
The reduced vDOS, $g(\omega)/\omega^2$, of the Debye theory is the
so-called Debye level $A_\mathrm{D}$~\cite{Kittel_1996,Ashcroft}.
$A_\mathrm{D}$ is calculated from the elastic moduli, $K$ and $G$, as
follows: $A_\mathrm{D} = 3/\omega_\mathrm{D}^3$, where
$\omega_\mathrm{D}$ is the Debye frequency defined as $\omega_\mathrm{D}
= \left[ 18 \pi^2 \rho/ ({2{c_\mathrm{T}}^{-3} + {c_\mathrm{L}}^{-3}})
\right]^{1/3}$, and $c_\mathrm{L} =\sqrt{(K+4G/3)/\rho}$ and
$c_\mathrm{T} = \sqrt{G/\rho}$ are the longitudinal and transverse sound
speeds, respectively.
Figure~\ref{fig5} presents the values of $\omega_\mathrm{D}$ and
$A_\mathrm{D}$ as functions of $\varepsilon_\mathrm{bend}$.
As the bulk modulus is approximately four times larger than the shear
modulus, $\omega_\mathrm{D}$ and $A_\mathrm{D}$ are mostly determined
with the shear modulus, i.e, $\omega_\mathrm{D} \approx \left( 9 \pi^2 \rho
\right)^{1/3} c_\mathrm{T}$ and $A_\mathrm{D} \approx 1/ \left(3 \pi^2 \rho
c_\mathrm{T}^3 \right)$.

As shown in Fig.~\ref{fig4}, the polymer glasses present clear excess
peaks over the Debye level, i.e., the BP.
The BP frequency, $\omega_\mathrm{BP}$, is defined as the frequency at
which $g(\omega)/\omega^2$ is maximal.
As $\varepsilon_\mathrm{bend}$ increases, $\omega_\mathrm{BP}$ shifts to
a higher frequency.
In addition, the height of the reduced vDOS,
$g(\omega_\mathrm{BP})/\omega_\mathrm{BP}^2$, becomes lower.
These shifts are observed in the region from
$\varepsilon_\mathrm{bend}=10$ to $10^3$ for $L=50$ and $3$.
Importantly, this region corresponds to the shear modulus $G$
variations, as shown in Figs.~\ref{fig3}c and~\ref{fig3}d.
As the bulk modulus is much larger than the shear modulus, the bulk
modulus should only have minor effects on the low-frequency spectra.
Therefore,  the BP of the proposed system should only be controlled by
the shear elasticity.

To confirm this hypothesis, the scaled vDOS
$g(\omega)/(\omega^2 A_\mathrm{D})$ is plotted as a function of the scaled frequency
$\omega/\omega_\mathrm{D}$ and presented in Fig.~\ref{fig6}.
As discussed above, $A_\mathrm{D}$ and $\omega_\mathrm{D}$ are
determined mostly by the shear modulus $G$.
Although deviations are observed for $L=50$, the scaled vDOSs collapse
for different values of $\varepsilon_\mathrm{bend}$.
In particular, an exact collapse is obtained for $L=3$.
This result indicates that the effects engendered by the bending
rigidity on the low-frequency spectra are comprised of the the shear
modulus changes.
A same collapse was observed in effects of pressure on the BP in the
covalent-bonding network glass~(Na$_2$FeSi$_3$O$_8$)~\cite{Monaco_2006}.
In addition, a previous experiment~\cite{Caponi_2011} demonstrated that
the effects of the polymerization are also comprised by the macroscopic
elasticity changes.
The collapsed results for (a) $L=50$ and (b) $L=3$ are consistent with
the experimental observation. 

According to the collapses observed in Fig.~\ref{fig6},
$\omega_\mathrm{BP} / \omega_\mathrm{D}$ does not depend on $\varepsilon_\mathrm{bend}$.
As stated previously, when $\varepsilon_\mathrm{bend}$ varies, $\omega_\mathrm{D}
\propto \rho^{1/3} c_\mathrm{T} \propto \rho^{-1/6} \sqrt{G}$.
As $\rho$ varies in a range of 15\%, as shown in Fig.~\ref{fig1}, the effect of
$\rho$ on $\omega_\mathrm{D}$ is weak. 
Thus, $\omega_\mathrm{D}$ is approximately proportional to $\sqrt{G}$, which leads to
$\omega_\mathrm{BP} \propto \sqrt{G}$ in the variation of
$\varepsilon_\mathrm{bend}$. 
The $\varepsilon_\mathrm{bend}$ dependence of the BP
frequency is determined by the shear modulus, which
is a macroscopic quantity describing the entire
system in an averaged manner.
It is noted that the recent study~\cite{Baggioli_2019} predicts
$\omega_\mathrm{BP} \propto \sqrt{G}$ from the phonon Green's function
with diffusive damping.
It might be interesting to study effects of $\varepsilon_\mathrm{bend}$
on phonon transport and the phonon's Green function.

According to the heterogeneous elasticity
theory~\cite{Schirmacher_2006,Schirmacher_2007,Schirmacher_2015}, the
spatial fluctuations of the local shear modulus $\delta G$ control nature of the BP
\footnote{
The value of $\delta G$ is quantified by the standard deviation of
probability distribution function of the local shear modulus~\cite{Mizuno_2013}.
}.
The collapse of $g(\omega)/( \omega^2 A_\mathrm{D})$ as a function of
$\omega/\omega_\mathrm{D}$ indicates that the shear modulus fluctuations
relative to the macroscopic value, $\delta G/G$, are constant for all
the cases of different bending rigidities.
Therefore, the results of this study can be explained as follows.
The increase in bending rigidity does not affect the shear modulus
fluctuations (relative to the macroscopic moduli) but only affects the
macroscopic shear modulus, which leads to the collapse of the scaled
vDOS.

\subsubsection{Participation ratio}
To further study the vibrational eigenstates, the participation ratio
$P^k$ that measures the extent of localization of the eigenmodes $k$ is
calculated as follows~\cite{Schober_1996,Mazzacurati_1996}:
\begin{equation}
P^k = \frac{1}{N_\mathrm{p}}\left[ \sum_{i=1}^{N_\mathrm{p}} (\boldsymbol{e}^k_i\cdot \boldsymbol{e}^k_i)^2\right]^{-1},
\end{equation}
where $\boldsymbol{e}^k_i$~$(i=1, 2, \cdots , N_\mathrm{p})$ are the
eigenvectors associated with the eigenfrequencies $\omega^k$~($i$ is the
index of the monomer particle and $N_\mathrm{p}$ is the number of monomer particles).
The $\boldsymbol{e}^k_i$ represents the displacements of each monomer
bead $i$ in the eigenmode $k$.
It is noted that $\boldsymbol{e}^k_i$ is obtained from the
diagonalization of the Hessian matrix and is orthonormalized as
$\sum_{i=1}^{N_\mathrm{p}} \boldsymbol{e}^k_i\cdot \boldsymbol{e}^l_i =
\delta_{kl}$~($\delta_{kl}$ is the Kronecker delta).
The following extreme cases can occur: ${P}^k = 2/3$ for an ideal
sinusoidal plane wave, ${P}^k = 1$ for an ideal mode in which all
constituent particles vibrate equally, and ${P}^k = 1/N_\mathrm{p} \ll 1$ for a
perfect localization,  which indicates that each vibrational state is
associated only with a single atom and that $e^k_i \cdot e^k_i=1$ for a
single $i$, otherwise $e^k_i \cdot e^k_i=0$.

Figure~\ref{fig7} presents the value of $P^k$ as a function of the
scaled frequency $\omega/\omega_\mathrm{D}$, for different
$\varepsilon_\mathrm{bend}$.
It is noted that the presented data are the binned average values.
Below the BP frequency $\omega_\mathrm{BP}$, $P^k$ progressively
decreases when $\omega$ decreases due to the spatially localized vibrations.
The low-frequency localization below $\omega_\mathrm{BP}$ has also been
observed in multiple
glasses~\cite{Schober_1996,Mazzacurati_1996,Taraskin_1997,Taraskin_1999}.
Importantly, $P^k$ below $\omega_\mathrm{BP}$ collapses between different values of $\varepsilon_\mathrm{bend}$.
This result indicates that the variations of not only the vDOS and the
vibrational states due to $\varepsilon_\mathrm{bend}$ can be
characterized by the macroscopic shear modulus changes.
However, $P^k$ does not collapse above $\omega_\mathrm{BP}$, as also shown in Fig.~\ref{fig7}.
This result is attributed to the fact that the high-frequency modes
above $\omega_\mathrm{BP}$ reflect microscopic vibrations that cannot
be captured by the macroscopic elasticity.

\subsubsection{Comparison with LJ glasses}
The low-frequency spectra are comparable to that of atomic LJ glasses reported in Ref.~\cite{Shimada:2018fp}.
As observed in Fig.~\ref{fig4}, the height of
$g(\omega_\mathrm{BP})/\omega_\mathrm{BP}^2$ of LJ glasses is higher
than that of polymer glasses, and $\omega_\mathrm{BP}$ is lower than
that of polymer glasses.
These observations are different from the study reported in
Ref.~\cite{Giuntoli:2018fv}, which demonstrated that the low-frequency
spectra of polymer glasses correspond to those atomic LJ glasses.
In Ref.~\cite{Giuntoli:2018fv}, the bonded monomers interact via a
harmonic potential with a large bonding energy scale of $k = 2500$.
This value is two orders of magnitude larger than
$\varepsilon_\mathrm{FENE} = 30$, investigated in this study.
With respect to the large bonding energy, the rigidity of the polymeric
chains has a smaller effect on the low-frequency spectra.
Therefore, the low-frequency spectra are mainly determined by the
non-bonding LJ interactions, whereas the elasticity is mainly determined
mainly by the bonding rigidity.
As a results, the BP decouples with the macroscopic elasticity, as
demonstrated in the previous study~\cite{Giuntoli:2018fv}.

In contrast to the the results presented in Ref.~\cite{Giuntoli:2018fv},
the rigidity of the polymeric chains is necessary to determine the
elasticity and the low-frequency spectra with respect to the bonding
energy scale of $\varepsilon_\mathrm{FENE} = 30$.
In fact, the $\varepsilon_\mathrm{bend}$ reduces the height of
$g(\omega_\mathrm{BP})/\omega_\mathrm{BP}^2$, as shown in
Fig~\ref{fig4}.
In this case, the BP couples with the macroscopic elasticity.
However, the plot of the scaled $g(\omega)/( \omega^2 A_\mathrm{D})$ as
a function of$\omega/\omega_\mathrm{D}$ does not collapse between the
polymer glasses and LJ glass, as shown in Fig.~\ref{fig6}.
The height of $g(\omega)/ (\omega^2 A_\mathrm{D})$ is consistent between
the polymer glasses and LJ glass, but $\omega/\omega_\mathrm{D}$ of the
LJ glass is lower than that of the polymer glasses.
This result indicates that vibrational states differences between
polymer glasses and LJ glasses cannot be described only by changes in
macroscopic elasticity, changes in the local elastic properties should
be considered as well~\cite{Monaco_2006,Niss_2007,Hong_2008,Mizuno2_2013}.

In addition, the length scale of collective vibrational modes in the BP
region is discussed.
For atomic LJ glasses, the length scale was evaluated as
$\xi_\mathrm{BP}=2 \pi c_\mathrm{T} /\omega_\mathrm{BP}$, which
corresponds to the size of approximately $23$ particle~\cite{Leonforte_2005}.
This length scale diverges near the isostatic point or the marginally
stable point,
theoretically~\cite{Wyart_2005,Wyart_2006,Wyart_2010,DeGiuli_2014} as
well as
numerically~\cite{Silbert_2005,Lerner_2014,Karimi_2015,Shimada_2018,Mizuno_2018}.
The present study evaluates the length scale of collective vibrational
modes in polymeric glasses as $\xi_\mathrm{BP}=2 \pi c_\mathrm{T}
/\omega_\mathrm{BP} \approx 12$, which corresponds to half of that for LJ glasses.
The vibrational modes in the BP region are more localized nature due to the polymerization.
Moreover, the value of $\xi_\mathrm{BP}$ is independent of the bending
rigidity $\varepsilon_\mathrm{bend}$ because of $\omega_\mathrm{BP}
\propto \omega_\mathrm{D} \propto c_\mathrm{T}$.
In other words, the bending rigidity does not affect the length scale of
the collective vibrational motions in the BP region.

\section{Discussion}
The glass transition temperature, elastic properties, and the
low-frequency vibrational spectra were studied in polymeric glasses.
In particular, the bending energy scale was highly varied for long
chains ($L=50$) and short chains ($L=3$).
As the system becomes rigid by increasing the bending rigidity, the
glass transition occurs at a higher temperature, leading to a lower
density in the glass phase.
The lowering density directly affects the isotropic bulk deformation,
but does not affect the shear elasticity.
The shear elasticity is controlled by only the bending rigidity only.
The non-affinity of polymeric glasses is much larger than that of atomic LJ glasses.
This is due to the more complex conformational relaxations of the
polymeric chains during non-affine deformation.
Even under an isotropic elastic deformation, the non-affine relaxation
process should be considered to describe the elastic response.

In addition, it is demonstrated that the BP frequency and its intensity
are simply scaled by the Debye frequency and the Debye level which are
mainly determined by the macroscopic shear modulus.
This result indicates that the BP is controlled by macroscopic shear
modulus and that the bending rigidity has a small impact on
heterogeneities of local elasticity properties.
The effects of the bending rigidity on the BP is similar to that of the
polymerization, which has also been explained by macroscopic elasticity
changes~\cite{Caponi_2011}.

The presented results provide a simple relationship between the BP and
the elasticity as well as the glass transition temperature.
As the system becomes more rigid by increasing the bending rigidity, the
glass transition temperature and the shear modulus are increased.
On the contrary, the bulk modulus $K$ decreases due to the decrease in
the density $\rho$ caused by the increase in the glass transition
temperature $T_g$.
However, the BP is mainly determined by the shear modulus $G$:
$\omega_\mathrm{BP} \propto \omega_\mathrm{D} \propto \sqrt{G}$.
Therefore, the glass transition temperature, the shear elasticity, and
the boson peak frequency are positively correlated.
A similar relationship between $T_g$ and $\omega_\mathrm{BP}$ was
observed experimentally in ionic liquids systems~\cite{Kofu_2015} and
also numerically in LJ glasses~\cite{Wang_2015}.
It is noted that the studies of Refs.~\cite{Kofu_2015,Wang_2015}
provided the relationship of $T_g \propto \omega_\mathrm{BP}^2$, but 
a clear power-law like relationship between $T_g$ and
$\omega_\mathrm{BP}$ was not observed in polymeric glasses.

Finally, it is worthwhile to discuss the structural relaxation in the
liquid state above the glass transition temperature.
A previous study~\cite{Larini_2008} has demonstrated the scaling
relationship between the structural relaxation time $\tau_\alpha$ and
the Debye-Waller factor $\langle u^2\rangle $ as $\tau_\alpha \propto
\exp\left( a \langle u^2\rangle ^{-1} + b \langle u^2\rangle ^{-2}
\right)$~(where $a, b$ are constants) for multiple glass-forming liquids
including polymeric glasses.
Here, the Debye-Waller factor in the harmonic
approximation~\cite{Shiba_2016} is estimated as $\langle u^2\rangle  =
3T \int_0^\infty {g(\omega)}/{\omega^2} d\omega \propto T
\omega_\mathrm{BP}^{-2} \propto T G^{-1}$.
It is naturally expected that the relaxation dynamics become drastically
slow by increasing the bending rigidity because of the following
relationship:
\begin{equation}
\tau_\alpha \propto \exp\left( \alpha \frac{\omega_\mathrm{BP}^{2}}{T} + \beta \frac{\omega_\mathrm{BP}^{4}}{T^2} \right) \propto \exp\left( \alpha' \frac{G}{T} + \beta' \frac{G^{2}}{T^2} \right),
\end{equation}
where $\alpha, \beta, \alpha', \beta'$ are constants.
This simple relationship demonstrates that the BP below $T_g$ and the
structural relaxation above $T_g$ are well correlated in the polymeric
glasses with varying the bending rigidity.
Further work is necessary to evaluate its validity by calculating $\tau_\alpha$.

\begin{acknowledgments}
The authors thank Atsushi Ikeda for useful discussions and suggestions.
This work was supported by JSPS KAKENHI Grant Numbers: 
 JP19K14670~(H.M.), JP17K14318~(T.M.), JP18H04476~(T.M.), JP18H01188~(K.K.),
 JP15K13550~(N.M.), and JP19H04206~(N.M.).
This work was also partially supported by the Asahi Glass Foundation and by the Post-K Supercomputing Project and the Elements Strategy Initiative for Catalysts and Batteries from the Ministry of Education, Culture, Sports, Science, and Technology.
The numerical calculations were performed at Research Center of Computational Science, Okazaki Research Facilities, National Institutes of Natural Sciences, Japan.
\end{acknowledgments}


%


\clearpage
\widetext

\setcounter{equation}{0}
\setcounter{figure}{0}
\setcounter{table}{0}
\setcounter{page}{1}

\noindent
{\Large{\textbf{Supplementary Material}}}
\vspace{5mm}
\begin{center}
\large{\textbf{Boson peak, elasticity, and glass transition temperature
 in polymer glasses:\\
 Effects of the rigidity of chain bending}}
\\
\vspace{2mm}
{Naoya Tomoshige, Hideyuki Mizuno, Tatsuya Mori, Kang Kim,
 and Nobuyuki Matubayasi}
\end{center}

\setcounter{section}{0}
\setcounter{equation}{0}
\renewcommand{\thesection}{S.\arabic{section}}
\renewcommand{\thesubsection}{S.\arabic{section} (\alph{subsection})}
\renewcommand{\theequation}{S.\arabic{equation}}
\renewcommand{\thefigure}{S\arabic{figure}}

\renewcommand{\bibnumfmt}[1]{[S#1]}
\renewcommand{\citenumfont}[1]{S#1}

\section{Formalism of the Hessian Matrix}

The Hessian matrix of the interaction potential $U(\bm{r})$ is generally
expressed as follows:
\begin{equation}
	\label{general_hessian}
	H_{nm}^{ab} = 
	\frac
	{\partial^2 U\left(\bm{r}\right)}
	{\partial r_n^a \partial r_m^b}
\quad (a, b=x, y, x)
\end{equation}
where $n$ and $m$ denote the particle number index ($n$, $m$=1, 2,
$\cdots$, $N_\mathrm{p}$).
As given in Ref.~\cite{Milkus:2018ha}, the following expressions are
useful using a generic argument $z$ for the first and second derivatives
of $U(z)$:
\begin{equation}
	\label{for_a_generic_argument_1}
	\frac
	{\partial U\left(z\right)}
	{\partial x}
	=
	\frac
	{\partial U\left(z\right)}
	{\partial z}
	\frac
	{\partial z}
	{\partial x},
\end{equation}
\begin{align}
	\label{for_a_generic_argument_2}
	\frac
	{\partial U^2\left(z\right)}
	{\partial x\partial y}
	=
	\frac
	{\partial U^2\left(z\right)}
	{\partial^2z}
	\frac
	{\partial z}
	{\partial x}
	\frac
	{\partial z}
	{\partial y}
	+
	\frac
	{\partial U\left(z\right)}
	{\partial z}
	\frac
	{\partial^2 z}
	{\partial x\partial y} 
=
	c\frac
	{\partial z}
	{\partial x}
	\frac
	{\partial z}
	{\partial y}
	+
	t\frac
	{\partial^2 z}
	{\partial x\partial y}.
\end{align}


\subsection{Two-body interaction}

For two-body interactions (FENE and LJ potentials), 
the distance between particles $i$ and $j$, $z = | \bm{r}_j - \bm{r}_i |
= r_{ij}$ is used and the following relationships are obtained:
\begin{equation}
	H_{nm}^{ab} = 
	\frac
	{\partial^2 U\left(r_{ij}\right)}
	{\partial r_n^a \partial r_m^b}
	=
	c_{ij}\frac
	{\partial r_{ij}}
	{\partial r_n^a}
	\frac
	{\partial r_{ij}}
	{\partial r_m^b}
	+
	t_{ij}
	\frac
	{\partial^2r_{ij}}
	{\partial r_n^a\partial r_m^b},
\end{equation}
with
\begin{equation}
	c_{ij} = \frac
	{\partial^2U\left(r_{ij}\right)}
	{\partial r_{ij}^2},\quad
	t_{ij} = \frac
	{\partial U\left(r_{ij}\right)}
	{\partial r_{ij}},
\end{equation}
and
\begin{equation}
	\frac{\partial r_{ij}}
	{\partial r_n^a}
	=
	\left(\delta_{nj}-\delta_{ni}\right)\hat n_{ij}^a,
\end{equation}
\begin{equation}
	\frac{\partial^2 r_{ij}}
	{\partial r_n^a\partial r_m^b}
	=
	\frac{1}{r_{ij}}
	\left(
	\delta_{nj}-\delta_{ni}
	\right)
	\left(
	\delta_{mj}-\delta_{mi}
	\right)
	\left(
	\delta_{ab}-\hat{n}_{ij}^a\hat{n}_{ij}^b
	\right),
\label{eq1:Milkus}
\end{equation}
where, $\hat n_{ij} = \bm{r}_{ij}/r_{ij}$ is the unit vector between the
particles $i$ and $j$.
These expressions are same as those presented in Ref.~\cite{Milkus:2018ha}.

\subsection{Three-body interaction}

For three-body interactions (bending potential), 
the bond angle of particles $i$, $j$, and $k$ is used as follows:
\begin{equation}
	z = \theta_{ijk} = \arccos\frac
	{\left( \bm{r}_j - \bm{r}_i \right) \cdot \left( \bm{r}_k - \bm{r}_i \right)}
	{r_{ij}r_{ki}}
	 = \arccos A_{ijk},
\end{equation}
hence,
\begin{equation}
	H_{nm}^{ab} = \frac
	{\partial^2 U\left(\theta_{ijk}\right)}
	{\partial r_n^a\partial r_m^b}
	=
	\tilde{c}_{ijk}\frac
	{\partial \theta_{ijk}}
	{\partial r_n^a}
	\frac
	{\partial \theta_{ijk}}
	{\partial r_m^b}
	+
	\tilde{t}_{ijk}
	\frac
	{\partial^2\theta_{ijk}}
	{\partial r_n^a\partial r_m^b}
\end{equation}
with
\begin{equation}
	\tilde{c}_{ijk} = \frac
	{\partial^2U\left(\theta_{ijk}\right)}
	{\partial \theta_{ijk}^2},
\quad
	\tilde{t}_{ijk} = \frac
	{\partial U\left(\theta_{ijk}\right)}
	{\partial \theta_{ijk}}.
\end{equation}
This following expression is obtained:
\begin{equation}
	H_{nm}^{ab} = 
	\frac
	{\tilde{c}_{ijk}}
	{\sin^2 \theta_{ijk}}
	\frac
	{\partial A_{ijk}}
	{\partial r_n^a}
	\frac
	{\partial A_{ijk}}
	{\partial r_m^b}
	-
	\frac
	{\tilde{t}_{ijk}}
	{\sin\theta_{ijk}}
	\left[
	\frac
	{\cos\theta_{ijk}}
	{\sin^2 \theta_{ijk}}
	\frac
	{\partial A_{ijk}}
	{\partial r_n^a}
	\frac
	{\partial A_{ijk}}
	{\partial r_m^b}
	+
	\frac
	{\partial^2 A_{ijk}}
	{\partial r_n^a\partial r_m^b}
	\right],
\label{eq2:Milkus}
\end{equation}
with
\begin{equation}
	\frac
	{\partial A_{ijk}}
	{\partial r_n^a}
	=
	\frac{1}{r_{ij}}
	\left(
	\delta_{nj}-\delta_{ni}
	\right)
	\left(
	\hat{n}_{ik}^a - \hat{n}_{ij}^a\cos\theta_{ijk}
	\right)
	+
	\frac{1}{r_{ik}}
	\left(
	\delta_{nk}-\delta_{ni}
	\right)
	\left(
	\hat{n}_{ij}^a - \hat{n}_{ik}^a\cos\theta_{ijk}
	\right),
\nonumber
\end{equation}
\begin{align}
	\frac
	{\partial^2 A_{ijk}}
	{\partial r_n^a\partial r_m^b}
	&=
	\frac
	{\delta_{ji}^n\delta_{ji}^m}{r_{ij}^2}
	\left[
	\left(
	3\hat{n}_{ij}^a\hat{n}_{ij}^b - \delta_{ab}
	\right)
	\cos\theta_{ijk}
	-
	\left(
	\hat{n}_{ik}^a\hat{n}_{ij}^b + \hat{n}_{ij}^a\hat{n}_{ik}^b
	\right)
	\right] &\nonumber\\
	&
\quad+
	\frac
	{\delta_{ji}^n\delta_{ki}^m}{r_{ij}r_{ik}}
	\left[
	\delta_{ab} + \hat{n}_{ij}^a\hat{n}_{ik}^b\cos\theta_{ijk}
	-
	\left(
	\hat{n}_{ik}^a\hat{n}_{ik}^b + \hat{n}_{ij}^a\hat{n}_{ij}^b
	\right)
	\right] &\nonumber\\
	&
\quad+
	\frac
	{\delta_{ki}^n\delta_{ji}^m}{r_{ij}r_{ik}}
	\left[
	\delta_{ab} + \hat{n}_{ik}^a\hat{n}_{ij}^b\cos\theta_{ijk}
	-
	\left(
	\hat{n}_{ik}^a\hat{n}_{ik}^b + \hat{n}_{ij}^a\hat{n}_{ij}^b
	\right)
	\right] &\nonumber\\
	&
\quad+
	\frac
	{\delta_{ki}^n\delta_{ki}^m}{r_{ik}^2}
	\left[
	\left(
	3\hat{n_{ik}^a}\hat{n_{ik}^b} - \delta_{ab}
	\right)
	\cos\theta_{ijk}
	-
	\left(
	\hat{n_{ij}^a}\hat{n_{ik}^b} + \hat{n_{ik}^a}\hat{n_{ij}^b}
	\right)
	\right].
\end{align}
The differences between the proposed calculation and the expression defined in Ref.~\cite{Milkus:2018ha}
arise from Eq.~(\ref{eq1:Milkus}) and the second term in the r.h.s. of Eq.~(\ref{eq2:Milkus}).
The overall profile of the vDOS $G(\omega)$ is not affected by implementing
the diagonalization of the Hessian matrix using the expressions in Ref.~\cite{Milkus:2018ha}.
A certain number of negative frequency
eigenmodes that have been reported in Ref.~\cite{Milkus:2018ha} have
also been observed.
On the contrary, the presented results of $g(\omega)$ using
Eqs.~(\ref{eq1:Milkus}) and (\ref{eq2:Milkus}) do not exhibit any negative eigenfrequency modes
(see Fig.~4 in the main text).

\end{document}